\documentclass[sigconf, conference, 10pt,letterpaper]{IEEEtran}

\usepackage{amsmath,amsfonts}
\usepackage{array}
\usepackage[caption=false,font=normalsize,labelfont=sf,textfont=sf]{subfig}
\usepackage{textcomp}
\usepackage{stfloats}
\usepackage{url}
\usepackage{verbatim}
\usepackage{graphicx}
\usepackage{cite}
\usepackage{xcolor}
\usepackage[linesnumbered,ruled,vlined]{algorithm2e}
\usepackage{multirow}
\usepackage{diagbox}

\newtheorem{definition}{Definition}

\begin{document}

\title{Execution time budget assignment for mixed criticality systems}

\author{
\IEEEauthorblockN{Mohamed Amine Khelassi, Yasmina Abdeddaïm}
\IEEEauthorblockA{\textit{Univ Gustave Eiffel, CNRS, LIGM},
F-77454 Marne-la-Vallée, France}
}



\maketitle

\begin{abstract}
In this paper we propose to quantify execution time variability of programs using statistical dispersion parameters. 
We show how the  execution time  variability can be exploited in  mixed criticality real-time systems. 
We propose a heuristic to compute the execution time budget to be allocated to each low criticality real-time task according to its execution time variability. We show using experiments and simulations
that the proposed heuristic reduces the probability of exceeding the allocated budget compared to algorithms which do not take into account the execution time variability parameter.  
\end{abstract}


\section{Introduction}


 In a mixed criticality system, programs of different criticality are executed  on the same processor. The challenge is that low criticality tasks do not disturb the good functioning of the high criticality ones.
 In real-time scheduling, since the original Vestal's model \cite{Vestal07}, a classical model has emerged, see  \cite{BurnsD17} for a complete survey.
In this model, tasks have several execution times budgets, one budget per  possible  criticality.  
If  a task does not signal its termination after the execution of its allocated budget at a certain criticality level, the system moves to the next criticality level. In every system criticality level, only tasks of criticality equal or higher to the criticality of the system  have to respect their deadlines. Therefore, low criticality tasks can be suspended to allow higher criticality tasks to meet their deadlines.
Much research in real-time mixed criticality scheduling assumes that the  execution time budgets in every criticality level are provided and focus on the challenge of maximizing the execution of low criticality tasks while guaranteeing that all high criticality tasks always meet their deadlines. The execution time budgets are often defined as estimates of the worst-case execution time at different certification requirements but few studies have focused on how to determine these budgets.

In this paper our aim is to determine the execution time budget of low criticality tasks so that all the tasks always respects their deadlines. If a job does not complete within its assigned budget, the job is suspended. We propose a heuristic to determine the execution time budgets, this heuristic uses an estimation of the variability of execution time of the task   to determine the budgets.
Many definitions have been proposed for the quantification of execution  time variability, in \cite{Variability_2001} and \cite{Flight_Control_Firmware} the execution  time variability is quantified as the ratio between run-time measurements and either the best, worst or mean execution time. 
In  \cite{OASIcs:2016:6899, AlainMerigot},  the execution time variability is considered to be the factor between the worst-case execution time computed in isolation and along with other workload. These approaches give a quantification that does not give information on how the execution times are distributed as we would like to have. 
In this work we  use statistical dispersion  parameters to quantify the execution time variability of a task. The idea is  to characterize the variability of execution times more accurately, so that we can use it to determine the budgets to be allocated to tasks.
Compared to probabilistic approaches \cite{DavisC19} that rely on the computation of the probabilistic response time to determine if the probability of a deadline miss respects a given threshold, our objective is  to use an estimate of the variability of execution times without using the complete distribution. Indeed, the methods that use the full distribution to compute the probabilistic response time have a high complexity for exact methods or have to make assumptions on the shape of the distributions for analytical methods.
The contributions of the paper are:
(1) We propose a  definition of execution time variability and a method for its quantification using statistical dispersion parameters,
(2) We propose a heuristic that uses the execution time variability to solve the scheduling problem of a mixed criticality system,
(3) We  evaluate our approach using simulations and benchmarks executed on an ARM-Cortex A53.


\section{Related work}

Concerning the use of statistical dispersion parameters in real-time systems, in \cite{Timing_Autonomous_Driving_Metrics, HPC_variability_CV} the authors use statistical dispersion parameters to measure the variability in the experimentation part to compare different execution time distributions of different modules of a system.  The difference with our proposition is that in our work statistical dispersion parameters are part of the task model and not a measure used to estimate the efficiency of an approach.
Decreasing the computation times of  LO-criticality tasks in order to avoid sacrificing all LO criticality tasks when the system is in high criticality  has been considered in \cite{Burns2013h,BaruahBG16,GuE16}. In these papers, a new model is used for mixed criticality systems  in which the budget of low criticality tasks is smaller when the system is in high criticality than their budget when the system is in low criticality. Our approach differs mainly in the following aspects. Firstly, in our model, all non-critical tasks have a single budget assigned to them. This budget has a practical significance and is linked to the distribution of task execution times. Secondly, our aim is to propose an approach that is agnostic to the scheduling algorithm used. Thirdly, the only monitoring we need to perform during execution is to stop jobs if they do not terminate at their assigned budget.

Our model  has some similarities with the approaches using probabilistic tasks models for mixed criticality systems   \cite{SinghSRBDG19,AbdeddaimM17,DraskovicAHT21} as we also consider that we have a distribution of execution times of the tasks. However, we don't use the distribution to compute the probability of deadlines misses, but we use it to guide us in the choice of budgets to assign to tasks so that deadlines are always met.

\section{Time variability quantification}

\begin{table}[]
    \centering
\begin{tabular}{|c|c|c|c|}
\hline
     & Merge sort & Quick sort & Insert sort \\
     \hline
   VWCET & 7,911 & 0,459  & 4,75 \\
   \hline
     sKw & 9,559 & -0,399 & -4,186 \\
  \hline
\end{tabular}
      \caption{VWCET and sKw of  Merge sort, Quick sort and Insert sort}
    \label{tab:dispertionparameters}
\end{table}
\begin{figure}
    \centering
  \begin{tabular}{c}
           \includegraphics[width=40mm]{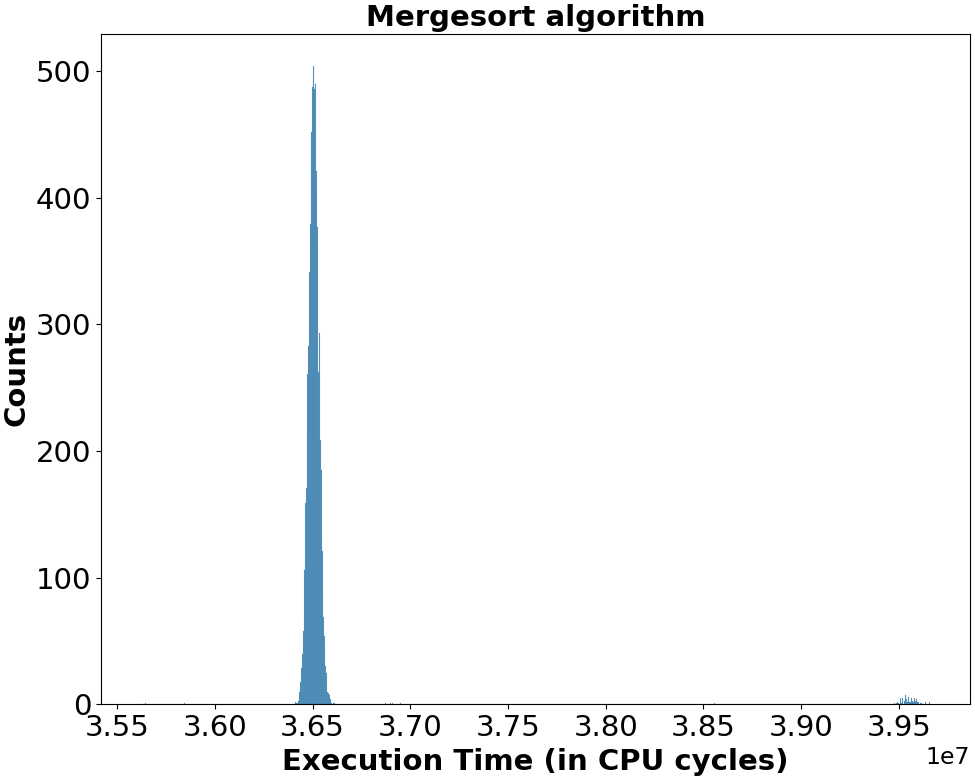}         \includegraphics[width=40mm]{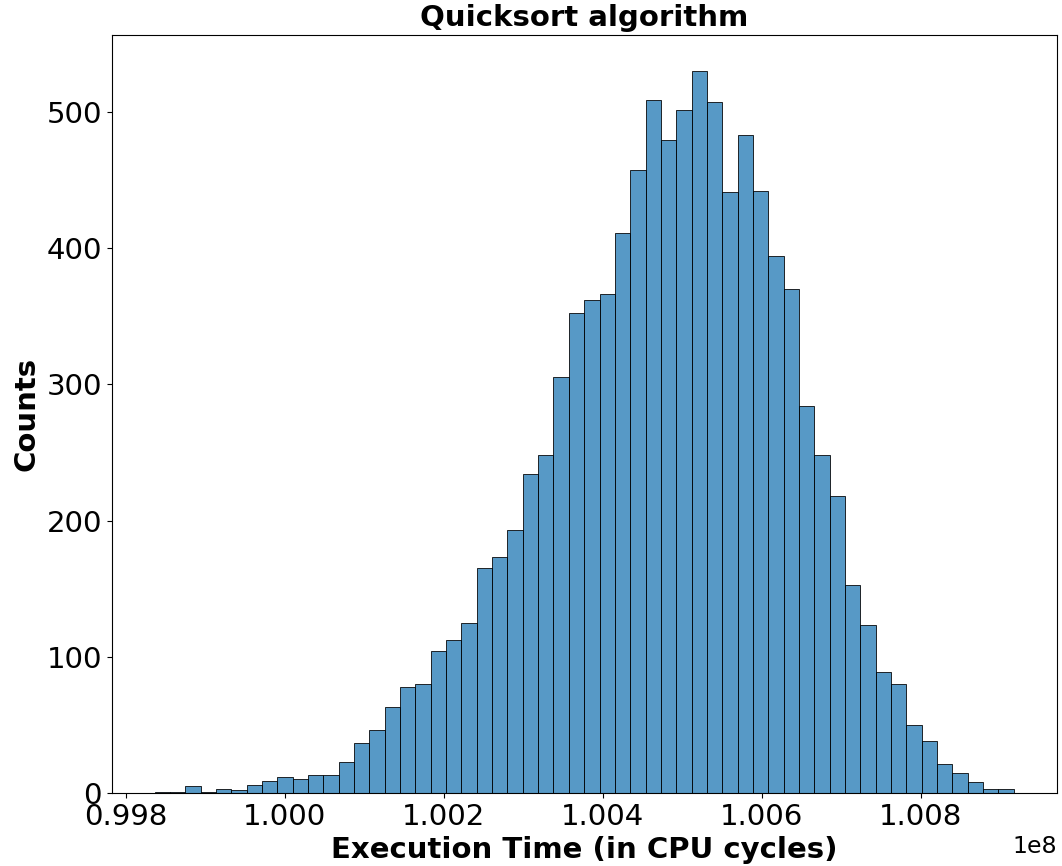}     \\   \includegraphics[width=40mm]{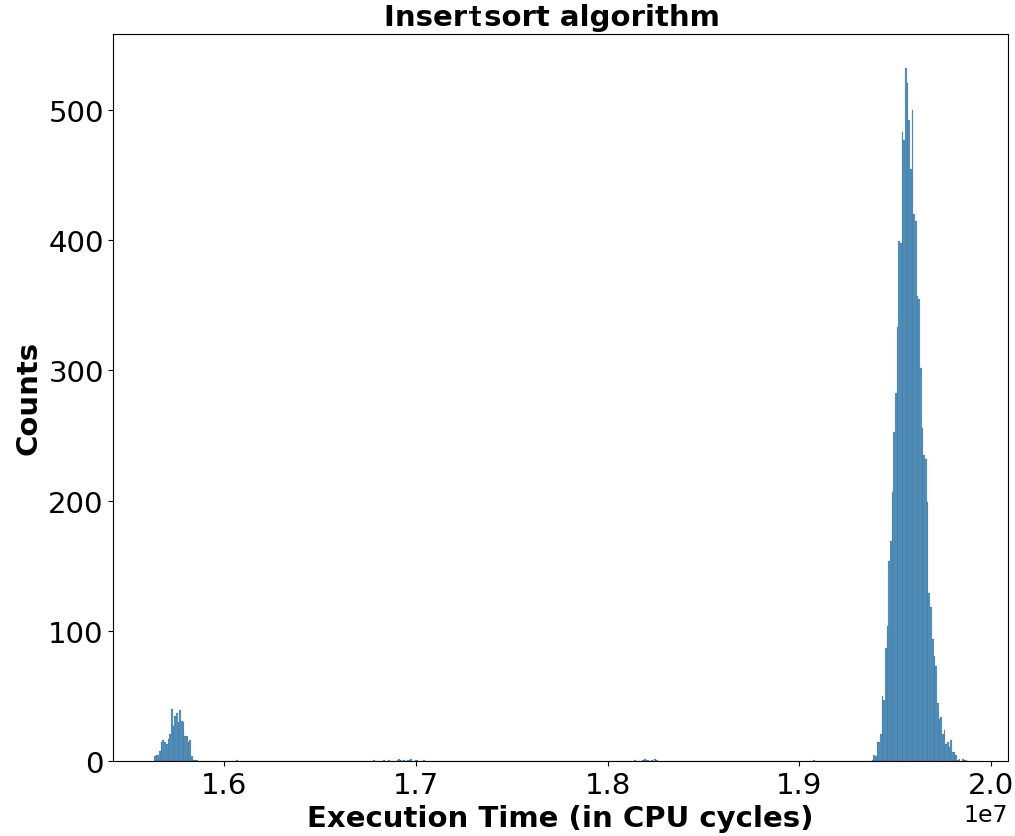} \\
             
        \end{tabular}
  
    \caption{Execution times (cycles) of programs executed in an ARM Cortex 53}
        \label{fig:variability}
\end{figure}

In this section we present our method to quantify the execution time variability of a program $P$.
Let $S(P)$ be a set of execution times of program $P$.
This set can be generated for example using repeated executions of the program in a specific configuration of the processor and/or using estimates of execution times for example in the worst and/or in the best case, however the way this set is computed is out of the scope of this paper. 
We note  $C_P$ the random variable taken its values in $S(P)$ with $p(C_P=x)$ being the probability of occurrence of $x$ in $S(P)$. Note that we do not make any assumption concerning the distribution function of this random variable.

We define the execution time variability as being quantified by a statistical dispersion parameter. In  Definition \ref{def:TV} we deliberately do not specify which statistical  dispersion parameter to use as we want to remain flexible in  the choice of the parameter to be used.
\begin{definition}[Execution time variability]
\label{def:TV}
The execution time variability of a program $P$ is defined by $TV$  a statistical dispersion parameter of $C_P$ the execution time random variable of program $P$.
\end{definition}
Dispersion parameters are used in statistics to  measure the tendency of the values of a distribution to be  scattered on either side of a value. For example the coefficient of variation measures the dispersion of a set of data around the mean. The higher the value of the coefficient of variation, the greater the dispersion of the values around the mean. The skewness  parameter \cite{Skewness} describes which side of the distribution has a longer tail. For unimodal distributions, symmetric  distributions should have a skewness value near zero, negative skewness values means that data are more shifted towards the maximum value, positive skewness values means that data are more shifted towards the minimum value of the distribution \cite{skwInterpret}.

In this work we propose our own dispersion parameter in Definition \ref{def:CVmax}. Our goal is to provide a  dispersion parameter that is not sensitive to the type of distribution the random variable follows.
This parameter is inspired from the classical coefficient of variation. In our case, it measures the dispersion of a set of data around the largest value instead of the mean. 

\begin{definition}[Coefficient of variation to the maximum]
\label{def:CVmax}
The coefficient of variation to the maximum $VWCET$ of $X$ a random variable with $n$ occurrences  $\{X_1, \ldots X_n\}$ is: 
\begin{equation}
\label{eq:CVmax}
    VWCET=\frac{\sqrt{\frac{\sum_{i=1}^{n}(X_i-M)^2}{n}}}{M}*100 \mbox{~with~} M = \underset{1 \leq i \leq n}{\max} X_i
\end{equation}
\end{definition}

Given that the VWCET is based on the sum of deviations between the WCET and the other data values, the smaller it is the less likely the data is far away from the WCET. 

To illustrate the VWCET and skewness parameters, Figure \ref{fig:variability} represents histograms of execution times (in terms of number of cycles) of three sorting programs, merge sort, quick sort and insert sort of the Mälardalen \cite{Mälardalen} benchmark. The execution times of the sorting algorithms were measured in a bare-metal environment using the Zynq UltraScale+ MPSoC ZCU104 Evaluation Kit. For every  algorithm we use the same input and  all the algorithms are executed in the same configuration of the processor i.e. the caches are disabled and one core executes the sorting program and  all other cores execute the same program. We can see that the shape of the histograms are are left-biased, centered, or right-biased. For every program we compute the VWCET and skewness statistical parameters. 

We can see in  Table \ref{tab:dispertionparameters} that the the $sKw$ value is close to zero for quick sort, negative for insert sort and positive for merge sort confirming the shapes of Figure \ref{fig:variability}. 
The $VWCET$ value of merge sort is the larger, then follows the $VWCET$ value of insert sort and the $VWCET$ quick sort is the smallest one. 

\section{Mixed criticality budget assignment problem}
\label{section:MC}

In this section we show  how execution time variability can be exploited in the real-time scheduling problem of mixed criticality systems. 
The goal is to define the execution time budgets to be assigned to tasks in such a way that: (1) the real-time task set is schedulable, i.e. all the tasks meet their deadlines whatever their criticality (2) critical tasks terminate at the latest at their allocated budget and (3) less critical tasks maximize the instances where they terminate the latest at their  allocated budget. We consider that if a job of a task does not terminate after the execution of its assigned budget, the job is stopped. In our model the jobs of the less critical tasks stopped only if during the execution they exceed their budget, otherwise they are executed and are guaranteed to meet their deadlines. 
We consider a task set $\Gamma$ of $n$  mixed criticality real-time tasks executed using a scheduling algorithm $Sched$. We do not specify whether the execution platform is uniprocessor or multiprocessor, however the scheduling algorithm must be sustainable with respect to task execution times, i.e. if the task set is schedulable using algorithm Sched, the task set is still schedulable using algorithm Sched if the task execution times are smaller.
Each task $\tau_i \in \Gamma$ is a tuple $(Budget_i,  TV_i, L_i,D_i,T_i )$ with:  
\begin{itemize}
    \item $Budget_i$: set of  possible budgets to be assigned to  $\tau_i$ with $p_i(b)$  is the probability that the budget $b$ is not exceeded
    \item $TV_i$:  execution time variability of  $\tau_i$
    \item $L_i\in\{LO, HI\}$: $\tau_i$ criticality, $LO$ less critical than $HI$
    \item $D_i \in {\mathbb{N^*}}$: relative deadline of  $\tau_i$
    \item $T_i \geq D_i $: task period i.e., its minimum inter-arrival time.
\end{itemize}

We consider that $TV_i$, the execution time variability, has been computed using given execution time random variables $C_i$. We suppose that the random variables $C_i$ of all the tasks are independent. The independence is necessary for the computation of the scores of Equation \ref{eq:score}.
The set $Budget_i$ is a subset of possible budgets of execution times selected from the possible values of $C_i$.  Each set of budgets of a task $\tau_i$ contains $m$ possible budgets noted $\{b_{1,i}, \ldots b_{m,i}\}$ ordered in a decreasing order of budgets with $b_{1,i}=WCET_i$ is the worst-case execution time of task $\tau_i$  with $p(b_{1,i})=1$.

Given a task set $\Gamma$ of $n$  mixed criticality real-time tasks, a  budget assignment for $\Gamma$ is a vector $B=(B_1, B_2, \ldots, B_n)$ with $\forall i \in 1 \ldots n, B_i \in Budget_i$ is the execution time budget assigned to task $\tau_i$. We define $\Gamma_B$ as the task set where every task $\tau_i$  is defined by $( B_i, L_i,D_i,T_i )$ with $B_i, L_i, D_i$ and $T_i$ are the execution time, the criticality, the deadline and the period of $\tau_i$ respectively.

Given a task set $\Gamma$ of $n$  mixed criticality real-time tasks  the score of  a budget  assignment $B=(B_1, B_2, \ldots, B_n)$ computes the probability that all the tasks do not exceed  their assigned budget and is defined as 
\begin{equation}
Score(B)=\prod_{i \in 1 \ldots n}{p_{i}(B_i)}
\label{eq:score}
\end{equation}
The score 
$Score(B_{LO})=\prod_{\substack{i \in 1 \ldots n\\ L_i \in LO}}{p_{i}(B_i)}$
is the score of the budget assignment of $LO$ criticality tasks and 
$Score(B_{HI})=\prod_{\substack{i \in 1 \ldots n\\ L_i \in HI}}{p_{i}(B_i)}$
is the score of the budget assignment of $HI$ criticality tasks.
\begin{definition}[Mixed criticality schedulability]
\label{def:score}
Given a task set $\Gamma$ of $n$  mixed criticality real-time tasks,  the task set $\Gamma$ is schedulable w.r.t. the budget assignment $B=(B_1, \ldots, B_n)$ and the scheduling algorithm Sched if and only if:
\begin{enumerate}
    \item The task set $\Gamma_B$ is schedulable according to the scheduling algorithm $Sched$
     \item $Score(B_{HI})=1$
\end{enumerate}
\end{definition}

\section{Budget assignment heuristic}
%


We consider the problem of finding the budget assignment of a task set $\Gamma$ such that $\Gamma$ is mixed criticality  schedulable and the probability of respecting the budget assignment for low criticality tasks is maximized. 
An optimal solution to the problem is to assign to all the high criticality tasks $\tau_i$ their largest value  as a budget and find the optimal assignment for  low criticality tasks among $m^{nbLO}$  possible assignments where $nbLO$ is the number of low criticality tasks. 
In this case the  schedulability test of the scheduling algorithm is executed in the worst-case $O(m^{nbLO})$ times. The more the number of low criticality tasks grows, the longer the computation time is, the latter is also impacted by the complexity of the used scheduling test.
For this reason, we propose a non optimal  greedy heuristic, Algorithm 1, where the  schedulability test is  executed in the worst-case $O(m \times {nbLO})$ times. In this algorithm the low criticality tasks are handled in a decreasing order of their execution time variability parameter.  The idea is to start by reducing the budget of the tasks that are the less  shifted towards their worst execution time. In Algorithm 1, first in  (lines 2-5) if the task set with the minimal assignment budget is not schedulable the problem is not schedulable. 

Otherwise a low criticality task with the highest execution time variability is selected. The budget for this task is decreased until the task set is schedulable (lines 8-14). If the task set is still not schedulable, the next task with the highest execution time variability parameter is selected.
\begin{algorithm}[!]
\label{alg:GeneralBudget}

\SetAlgoLined         
\SetKwInOut{KwIn}{Input}                
\SetKwInOut{KwOut}{Output}            
\KwIn{$\Gamma = \{\tau_i, i \in 1..n\}$}
\KwOut{system is not schedulable or $B = (B_1, \ldots B_n)$}
\BlankLine

$Set =$ set of all low criticality tasks

$\forall \tau_i,$ if $\tau_i \in Set,  B_i = b_{m,i}$ else  $B_i = b_{1,i}$\;

\If{$\Gamma_B$ not schedulable using algorithm Sched}{
return system is not schedulable\ 
}

$\forall \tau_i \in Set, B_i = b_{1,i}$

\While{$\Gamma_B$ not schedulable using algorithm Sched} {

$i$ is the index of $\tau_i$ with $\forall \tau_j \in Set, TV_i \geq TV_j$

\For{$r \in 2..m$}{
$B_i =  b_{r,i},$

\If{$\Gamma_B$ is  schedulable using algorithm Sched}{
return $B = (B_1, \ldots B_n)$\;
}
}

$\tau_i$ is removed from  $Set$\;

\If{$Set = \emptyset$}{
return system is not schedulable \;
}

}
return $B = (B_1, \ldots B_n)$\;
\caption{Time variability heuristic}
\end{algorithm}

We illustrate our budget assignment algorithm using the following example.

Let $\tau_1$, $\tau_2$ be two LO criticality tasks and $\tau_3$ a HI criticality task scheduled in a mono-processor according to rate monotonic scheduling algorithm. For every task $\tau_i$, $S(\tau_i)$ is a set of $100$ executions times of $\tau_i$ with:
\begin{itemize}
    \item $S(\tau_1)$ is composed of $10$ occurrences of value $1$,  $20$ occurrences of value $2$ and  $70$ occurrences of value $3$.
     \item $S(\tau_2)$ is composed of $40$ occurrences of value $1$,  $50$ occurrences of value $2$ and  $10$ occurrences of value $3$.
      \item $S(\tau_3)$ is composed of $10$ occurrences of value $1$,  $10$ occurrences of value $2$ and  $80$ occurrences of value $3$.
\end{itemize}

The VWCET of $\tau_1$ and $\tau_2$ computed using Formula \ref{eq:CVmax} are respectively $VWCET_1=0.258$, $VWCET_2=0.48$. We don't compute  $VWCET_3$ because the assigned budget of HI criticality task do not depend on the variability of the task.

Let  $\Gamma$ be a mixed criticality task set, as defined in Section \ref{section:MC}, composed by $\tau_1=((1,2,3), 0.258, LO, 6, 6)$, $\tau_2=((1,2,3), 0.48, LO, 9, 9)$ and $\tau_3=((1,2,3), VWCET_3, HI, 12, 12)$.

Using response time analysis \cite{RTA} the worst-case probability deadline miss of  $\tau_3$   is $0.204$ which is quite high for a high criticality task. We need to reduce LO criticality task execution times to ensure the schedulability.
Algorithm 1 applied to $\Gamma$ first assigns to all the tasks  the budget $3$. Then the task $\tau_2$ is selected as it is the task with the largest time variability among the set of low criticality tasks. The budget of $\tau_2$ is decreased until the budget $1$ as the task set is schedulable with this budget.
Algorithm 1 returns the set of budgets $B_1=3$, $B_2=1$ and $B_3=3$ for the tasks $\tau_1$, $\tau_2$ and $\tau_3$ respectively. The score $Score(B_{HI})=1$ because the probability of not exceeding the budget $B_3=3$ is equal to $1$ and $Score(B_{LO})=1 \times 0.4= 0.4$  with $1$ is the probability of   not exceeding the budget $B_1=3$ and $0.4$ is the probability of   not exceeding the budget $B_2=1$. Note that the budgets computed with Algorithm 1 are the budgets with the optimal score.

\section{Simulation-based evaluation}
In this section, we present simulation-based experiments to evaluate the proposed heuristic.
We consider uniprocessor earliest deadline first algorithm. We can consider without loss of generality  that all the tasks are of low criticality because all the algorithms we are comparing assign to the  high criticality tasks WCET as budget.
For each task we generate a maximal random utilization that ranges from 1 to 1.45 using \cite{RobTaskSetGen}. We generate periods  using a uniform distribution in the range $[4,102]$ and a deadline using a uniform distribution in the range $[T_i/2,T_i]$. We generate the minimal random utilizations by reducing the maximal utilizations by a value that ranges between $1$ and $45$ percent.
We use a truncated normal distribution, where the mean is a random (uniform) value ranging from the Best case execution time BCET to the WCET of the task, and the standard deviation is a random value proportional to the span of execution times $(WCET-BCET)/(x)$ where $x$ follows a uniform distribution in the range $[2,40]$. 
We generate $1000$ task sets composed of $6$ tasks. For each task $\tau_i$, $Budget_i=\{WCET_i, b_1, b_2, Median_i\}$ where $b_1$ and $b_2$ are the $80th$ and $60th$ percentiles respectively and $Median_i$ is the median value. 
We discard task sets with $\sum_{i=1\ldots n}\frac{BCET_i}{T_i}> 1$, and  task sets which do not produce a solution with at least one algorithm in the list of the algorithms that we compare. 

We generate three categories of task sets:
\begin{itemize}
    \item Scenario  1: the first 80\% of the task distributions have a skewness value greater than 2, the next 10\% have a skewness value between 2 and -2, and the final 10\% have a skewness value less than -2.
    \item Scenario 2: the first 10\% of the task distributions have a skewness value greater than 2, the next 10\% have a skewness value between 2 and -2, and the final 80\% have a skewness value less than -2.
    \item Scenario 3: distributions are generated randomly.
\end{itemize}

We test Algorithm 1 in two cases: 
\begin{itemize}
        \item VWCET: Algorithm 1 with $\forall \tau_i$,  $TV_i=VWCET(C_i)$,
    \item  sKw: Algorithm 1 with $\forall \tau_i$,  $TV_i$ is the skewness of $C_i$.
\end{itemize}

We compare Algorithm 1 to the following algorithms  that don't consider variability parameter as a criteria to select the task to reduce it's time budget:
\begin{itemize}
    \item Medians: low tasks are assigned the median as a budget.
    \item Opt: optimal solution by testing all the configurations and applying the schedulability test  $O(m^{nbLO})$ times.
    \item Periods: similar to Algorithm 1 but tasks are selected in an ascending order of periods in line 8.
      \item Deadlines:  similar to Algorithm 1 but tasks are selected in an ascending order of deadlines in line 8.
    \item Random:  similar to Algorithm 1 but tasks are selected randomly in line 8.

\end{itemize}

\begin{figure}[t]
     \centering
    
         \centering
         \includegraphics[width=6.5cm, height=5.5cm]{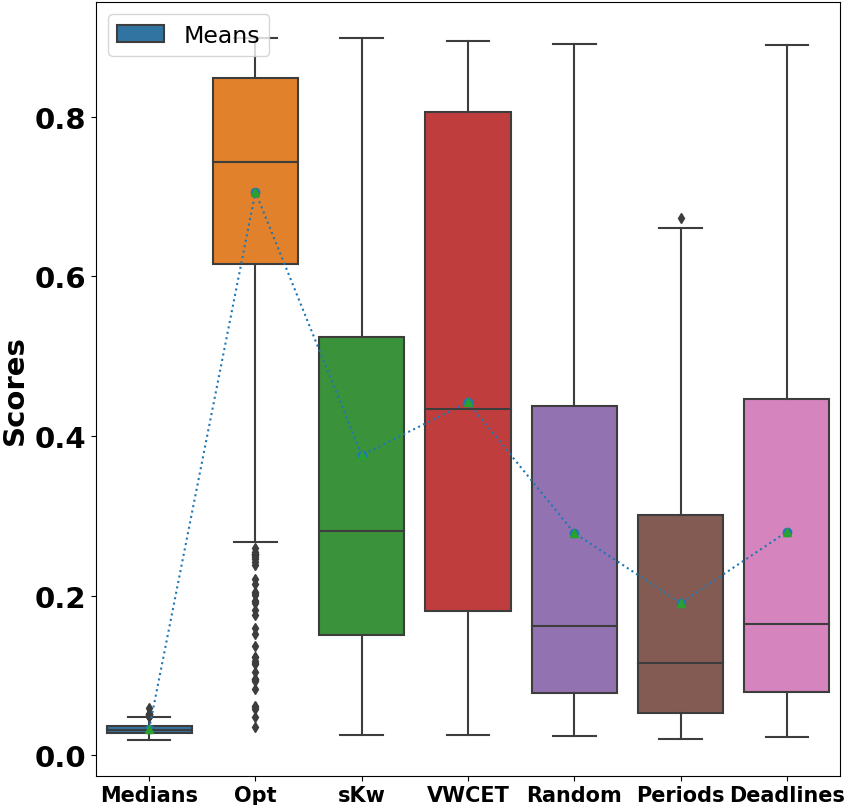}
          \caption{Algorithms comparison for Scenario 1}
         \label{fig:y equals x}

         \label{fig:Distribution2}
\end{figure}

\begin{figure}[t]
     \centering

         \includegraphics[width=6.5cm, height=5.5cm]{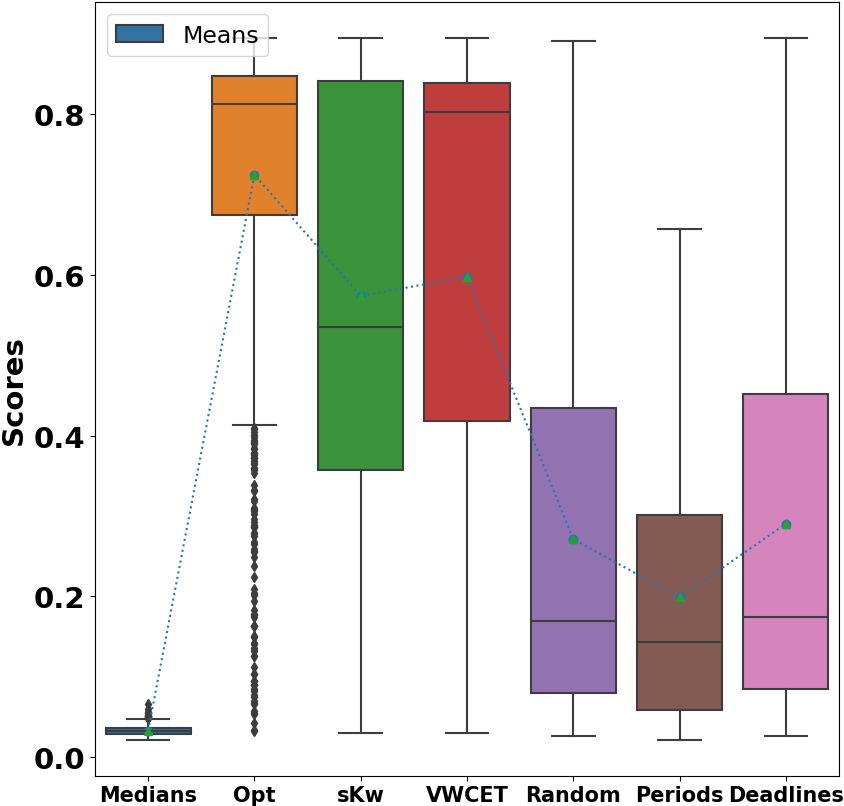}
          \caption{Algorithms comparison for Scenario 2}

         \label{fig:Distribution4}
\end{figure}

\begin{figure}[t]
    \centering   
    \includegraphics[width=6.5cm, height=5.5cm]{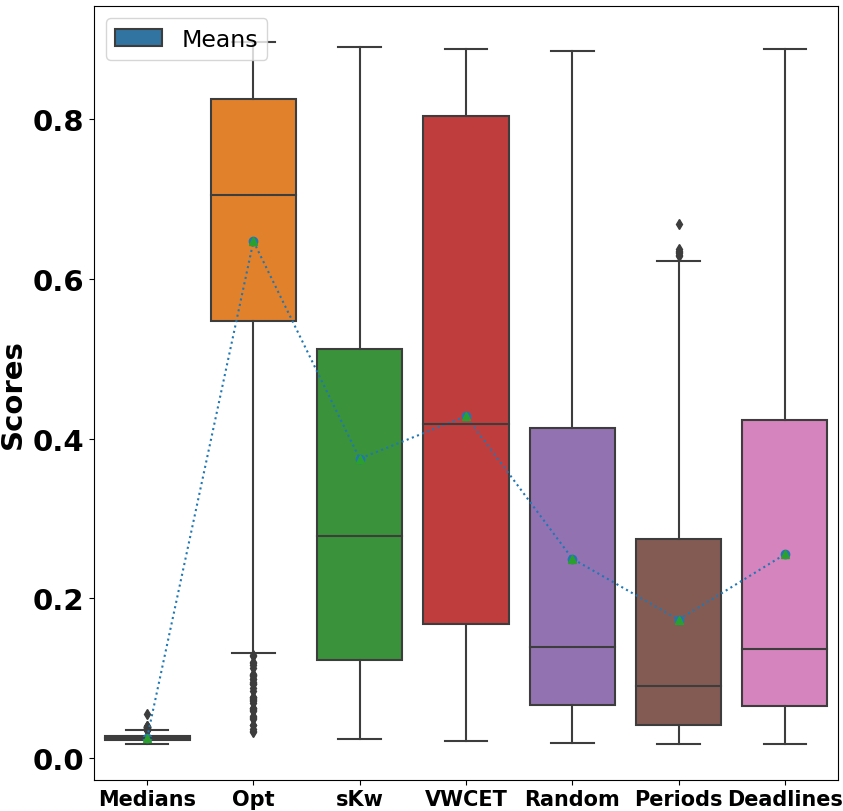}

 \caption{Algorithms comparison for Scenario 3}
       \label{fig:Distribution5}
\end{figure}

Figures \ref{fig:Distribution2}, \ref{fig:Distribution4} and  \ref{fig:Distribution5} depict the scores obtained  by the different algorithms in the scenarios 1, 2 and 3 respectively suing box plot representation. The mean score is also represented in each figure. We conclude that:
\begin{enumerate}
    \item  Assigning all the budgets of  all the tasks to their median value budget result in a score close to $0$,
    \item Using the execution time variability gives better score than using the  periods, or deadlines or a random,
    \item VWCET  gives better scores than the skewness parameter
   \item  The difference between the scores of the optimal solution and the ones obtained by the heuristics have the same order of magnitude in all the scenarios. This suggests that our approach is not highly affected by the shape of execution times distribution. 
\end{enumerate}

\begin{figure}
    \centering
    \includegraphics[width=6.5cm, height=5.5cm]{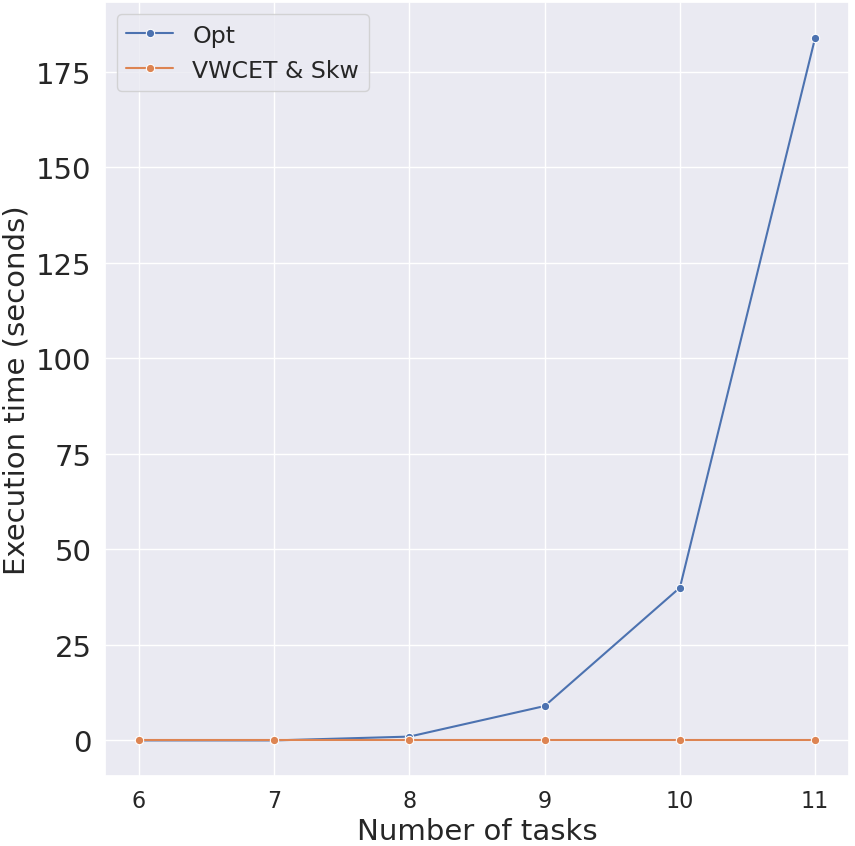}
    \caption{Computation time of Opt compared to Algorithm 1}
    \label{fig:Opt_vs_VWCET}
\end{figure}

In Figure \ref{fig:Opt_vs_VWCET}, we compare the time needed to compute the set of assigned budgets of the optimal algorithm Opt and the Algorithm 1. We can see that from task sets with $8$ tasks, the execution time of Opt increases exponentially.

\section{Benchmark-based evaluation}

We performed our experiments on the Zynq UltraScale+ MPSoC ZCU104 platform, featuring a quad-core Arm Cortex A53 processor. We use FreeRTOS \cite{FreeRTOS}, an open source, small, scalable and well documented real-time OS, along with ESFree \cite{Kase2016EfficientSL} as the scheduling library in order to incorporate advanced features that are not nativaly proposed by FreeRTOS, such as timing error detection and handling mechanisms for scenarios involving task time budget excess and deadline misses. ESFree also provides an extended version of the FreeRTOS task control block (TCB) that includes additional information for managing periodic tasks such as current execution time, WCET and deadline of the task. 
FreeRTOS measures time using a tick count variable. The RTOS tick interrupt increments the tick count with strict temporal accuracy allowing the kernel to measure time to a resolution of the chosen timer interrupt frequency. In our case we choose  a resolution of $1000$ Hz, this translates into a time resolution of $1$ ms, consequently execution time smaller than $1$ ms cannot be measured. 
The tick interrupt calls an application defined hook  function where we check which task is currently under execution and increase it's execution time variable by $1$ ms.
We generate a set of execution times by executing $6$ programs from TACLeBench \cite{tacle} and Mälardalen \cite{Mälardalen} benchmarks under the following configuration:
(1) we consider asymmetric multiprocessing with FreeRTOS, meaning each core of a multicore processor runs its own independent instance of FreeRTOS
(2) we do not vary the inputs of the programs
(3) caches are disabled 
(4) cores 1, 2  and 3 execute \cite{CIG} a program that iterates over a large array 
(5) core 4 executes the $6$ programs and perform measurements to collect a set of execution times
(6) the scheduling algorithm is  preemptive rate monotonic.
\begin{table}[]
\centering
\begin{tabular}{|c|c|c|c|c|c|c|}
\hline
Prog  & epic & fft & iSort & ludcmp & matrix & sha \\ \hline
T(ms) & 450  & 430 & 70    & 440   & 677    & 460 \\ \hline
D(ms) & 152  & 18  & 10    & 52    & 390    & 247 \\ \hline
\end{tabular}
\caption{Program's parameters}
\label{tab:prog-info}
\end{table}

Table \ref{tab:prog-info} gives for every program its deadline and its period. All the programs are considered of LO criticality.
For each program $i$,  the   budgets  are $Budget_i=\{WCET_i, b_1, ... , b_7, Median_i\}$ where $b_1$, ... , $b_7$ are the $99th$, $97th$, $95th$, $90th$, $80th$, $70th$, $60th$ percentiles respectively and $Median_i$ is the median value.

Using the generated execution times we apply:
\begin{itemize}
        \item VWCET:  Algorithms 1 with $\forall \tau_i$,  $TV_i=VWCET(C_i)$,
    \item  sKw: Algorithms 1 with $\forall \tau_i$,  $TV_i$ is the skewness of $C_i$.
\end{itemize}

A response time analysis \cite{RTA} is used as schedulability test.
We compare Algorithm 1 to the same algorithms used in the simulation-based evaluation i.e. Medians, Opt, Periods, Deadlines and Random.
Table \ref{tab:scoresESFree} depicts the scores obtained for every program using every algorithm. The score $\Pi$ is the score computed using Formula 2. 
\begin{table}[]
\begin{tabular}{|c|c|c|c|cl|cl|cl|}
\hline
Prog   & Medians & Random & Periods & \multicolumn{2}{c|}{sKw}  & \multicolumn{2}{c|}{VWCET} & \multicolumn{2}{c|}{Opt}  \\ \hline
epic   & 0.51    & 0.73   & 0.73    & \multicolumn{2}{c|}{1}    & \multicolumn{2}{c|}{1}     & \multicolumn{2}{c|}{0.97} \\ \hline
fft    & 0.50    & 0.99   & 0.71    & \multicolumn{2}{c|}{0.80} & \multicolumn{2}{c|}{0.71}  & \multicolumn{2}{c|}{0.99} \\ \hline
iSort  & 1       & 1      & 1       & \multicolumn{2}{c|}{1}    & \multicolumn{2}{c|}{1}     & \multicolumn{2}{c|}{1}    \\ \hline
ludcmp  & 0.51    & 0.72   & 0.72    & \multicolumn{2}{c|}{0.72} & \multicolumn{2}{c|}{0.72}  & \multicolumn{2}{c|}{0.98} \\ \hline
matrix & 0.50    & 0.72   & 1       & \multicolumn{2}{c|}{1}    & \multicolumn{2}{c|}{0.80}  & \multicolumn{2}{c|}{0.90} \\ \hline
sha    & 0.53    & 0.72   & 0.99    & \multicolumn{2}{c|}{0.72} & \multicolumn{2}{c|}{1}     & \multicolumn{2}{c|}{0.97} \\ \hline
$\Pi$  & 0.03    & 0.27   & 0.37    & \multicolumn{2}{c|}{0.42} & \multicolumn{2}{c|}{0.41}  & \multicolumn{2}{c|}{0.83} \\ \hline
\end{tabular}
\caption{Algorithms Scores}
\label{tab:scoresESFree}
\end{table}
We can see that Algorithm 1 using variability parameters sKw and VWCET yields better results than the other algorithms Medians, Random and Periods.
When comparing the optimal solution computed using algorithm Opt to the proposed  heuristic that uses time variability we can see that the optimal solution in order to maximize the overall scores of the tasks reduces the time budget of 5 of the 6 tasks, in difference to the proposed  heuristic where 3 of the 6 tasks will always meet their time budget. Using algorithm Periods only 2 of the 6 tasks will always meet their time budget. The greater the number of tasks whose jobs always meet the assigned budget, the fewer tasks we need to monitor during execution.
\begin{table}[]
\centering
\begin{tabular}{|l|c|c|c|c|c|c|}
\hline
Program $i$ & epic & fft  & iSort & ludcmp & matrix & sha  \\ \hline
$p(B_i) $   (VWCET)   & 1    & 0.71 & 1     & 0.72  & 0.80   & 1    \\ \hline
$ratio_i$  (VWCET)    & 1    & 0.69 & 1     & 0.75  & 0.82   & 0.99 \\ \hline
    \hline
$p(B_i) $  (sKw) & 1    & 0.80 & 1     & 0.72   & 1      & 0.72 \\ \hline
$ratio_i$    (sKw)       & 0.99 & 0.84 & 1     & 0.75   & 1      & 0.72 \\ \hline
\end{tabular}
\caption{The probabilities of meeting the execution time budget}
\label{tab:ESFreeInAction}
\end{table}
Using FreeRTOS OS and the library ESFree, we executed the $6$ programs during 10 minutes using rate monotonic preemptive scheduling algorithm.  If a job  do not terminate at its assigned  budget computed using Algorithm 1, the job is stopped. We count for every task $\tau_i$  the number of stopped jobs and computes $ratio_i$ the ratio between the number of stopped jobs and the number of activated jobs of task $\tau_i$.
Table \ref{tab:ESFreeInAction} presents the comparison of the  probability of meeting the execution time budget per task to the ratios obtained. For VWCET the observed task performance aligns well and sometimes slightly exceeds the theoretically computed time budgets, this conclusion holds true for skewness as well. We can deduce that the overhead induced when stopping a job is not high.

\section{Conclusion}

This paper introduces an approach for the scheduling problem of mixed criticality systems.
We show that the time variability of execution times can be used to compute the execution time budget to be allocated to low-criticality tasks in order to prevent these tasks from disrupting the functioning of high-criticality tasks, while ensuring that low-criticality tasks are not always stopped.
We show that the proposed heuristic outperforms the budget selection strategies that are agnostic to the shape of the distribution of execution time budgets.
In the future, we plan to apply our approach in an online algorithm where task budgets are adjusted during the execution.

\bibliographystyle{IEEEtran}
\bibliography{referencesbib}

\end{document}